\documentclass[aps,prd,preprint,groupedaddress]{revtex4-2}

\usepackage{graphicx}
\usepackage{dcolumn}
\usepackage{bm}

\usepackage[utf8]{inputenc}
\usepackage[T1]{fontenc}
\usepackage{mathptmx}
\usepackage{etoolbox}

\usepackage{amssymb,amsmath,placeins,epsfig,array}
\usepackage[usenames,dvipsnames]{color}
\usepackage[normalem]{ulem}
\usepackage[breaklinks,colorlinks,urlcolor=blue,citecolor=blue,linkcolor=blue]{hyperref}
\usepackage{lineno}
\usepackage{graphicx}
\usepackage{float}
\usepackage{mathptmx}

\usepackage{scalerel}
\usepackage{tikz}
\usetikzlibrary{svg.path}

\definecolor{orcidlogocol}{HTML}{A6CE39}
\tikzset{
  orcidlogo/.pic={
    \fill[orcidlogocol] svg{M256,128c0,70.7-57.3,128-128,128C57.3,256,0,198.7,0,128C0,57.3,57.3,0,128,0C198.7,0,256,57.3,256,128z};
    \fill[white] svg{M86.3,186.2H70.9V79.1h15.4v48.4V186.2z}
                 svg{M108.9,79.1h41.6c39.6,0,57,28.3,57,53.6c0,27.5-21.5,53.6-56.8,53.6h-41.8V79.1z M124.3,172.4h24.5c34.9,0,42.9-26.5,42.9-39.7c0-21.5-13.7-39.7-43.7-39.7h-23.7V172.4z}
                 svg{M88.7,56.8c0,5.5-4.5,10.1-10.1,10.1c-5.6,0-10.1-4.6-10.1-10.1c0-5.6,4.5-10.1,10.1-10.1C84.2,46.7,88.7,51.3,88.7,56.8z};
  }
}

\newcommand\orcidicon[1]{\href{https://orcid.org/#1}{\mbox{\scalerel*{
\begin{tikzpicture}[yscale=-1,transform shape]
\pic{orcidlogo};
\end{tikzpicture}
}{|}}}}

\usepackage{hyperref}
\usepackage{graphicx}
\usepackage{dcolumn}
\usepackage{bm}
\usepackage{subfigure}
\usepackage{epstopdf}
\usepackage{hyperref}
\usepackage{color}
\usepackage[utf8]{inputenc}

\makeatletter
\def\@email#1#2{%
 \endgroup
 \patchcmd{\titleblock@produce}
  {\frontmatter@RRAPformat}
  {\frontmatter@RRAPformat{\produce@RRAP{*#1\href{mailto:#2}{#2}}}\frontmatter@RRAPformat}
  {}{}
}%

\makeatother

\newcommand\CWM{William \& Mary, Williamsburg, VA 23185}
\newcommand\JLab{Thomas Jefferson National Accelerator Facility, Newport News, VA 23606}

\begin{document}
\title{Polarized internal target experiments based on EIC beams}
\author{B.~Wojtsekhowski}
\email{bogdanw@jlab.org}
\affiliation{\CWM} \affiliation{\JLab} 
\begin{abstract}The Electron-Ion Collider is under construction at BNL. It will have high-energy high-intensity polarized beams of
electrons and hadrons.
These beams will allow a high accuracy investigation of nucleon structure in the low- to very-low-x DIS regime.
At the same time, similar to the realization at HERA, these beams could be used with an internal target for a very productive 
investigation of medium- to high-x nucleon structure.
Due to a novel regime of electron beam operation and its high polarization and intensity, the Figure-of-Merit
of an internal target experiment at EIC will be 500+ times higher than was obtained by HERMES.
\end{abstract}

\maketitle

\section{Introduction}

Experiments with an electron beam and an internal target in the storage rings were proposed a long time ago, see e.g. Ref.~\cite{VEP-1}.
They are productive due to fast damping of the beam particle oscillations thanks to intense synchrotron radiation. 
The important features of the internal target method were formulated soon after~\cite{Budker-67, Zelevinski-69, Belyaev-70}.
These include a continuous electron beam on the target, high luminosity for data taking, reduced interaction of the produced
particles in the target, the possibility of using polarized atoms as a target, and the use of unique beams of positrons and antiprotons.
Internal targets were used at BINP to study the light nuclei and a polarized deuteron~\cite{Popov-82, Dmitriev-85, Popov-87, NEP, Gilman-90}.

In the 1980s, with the construction of the multi GeV electron storage rings at SLAC and DESY, two large scale experiments were 
proposed~\cite{Milner-84, VANBIBBER}.
They were motivated by the discovery of a nucleon ``spin puzzle" by the EMC collaboration and 
the wide physics program on the meson and hadron structure which could be realized.
The proposal for HERA~\cite{McKeown-87, Milner-87}, focused on spin structure functions, was accepted and led to
many key advances in hadron physics~\cite{Milner-2021}.

\begin{figure}[ht!]
	\centering
	\includegraphics[trim = 0 0 0 0, width=0.7\columnwidth, angle =-0.5]{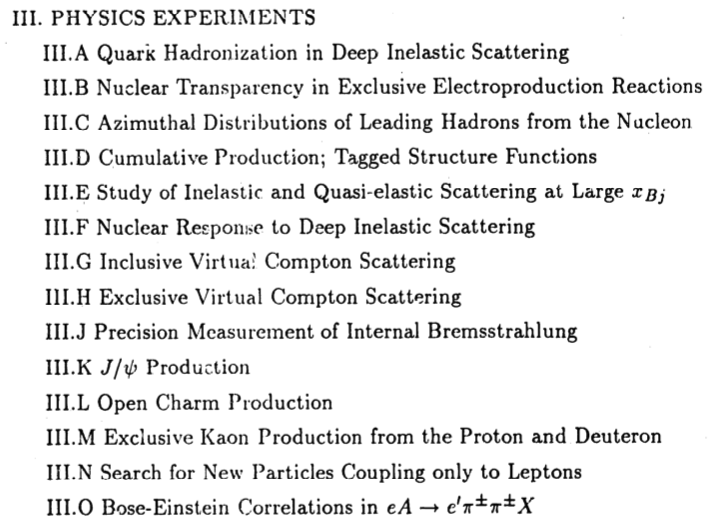}
	\caption{Physics program of PEGASYS~\cite{PEGASYS}.}
\label{fig:pegasys}
\end{figure}

The physics program on PEGASYS included 14 topics shown in Fig.~\ref{fig:pegasys} taken from Ref.~\cite{PEGASYS}.
After experiments at HERMES, SLAC, Compass and JLab, most of these topics were advanced very significantly.
Several of them, e.g. $J/\psi$ production and Tagged structure functions, have become hot topics in recent years.
At the same time, as is typical in science, with higher experimental productivity the next level of understanding of the hadron nature could be achieved.

The EIC Yellow Report~\cite{EIC-Yellow} briefly indicated ``opportunities from fixed target mode operation".
The experiment productivity or Figure-of-Merit is defined by a product of the beam intensity, the square of the beam polarization, the target thickness (nucleon/cm$^2$), and the square of the nucleon polarization (including dilution factor): FOM $\,=\, I_e \times P_{e}^2 \times t_{_N} \times P_{_N}^2$.
We show here that the EIC electron beam allows a huge increase in the experimental Figure-of-Merit, see also Ref.~\cite{BW-23}.
The EIC electron beam also allows us to create a tagged photon beam with energy up to 18 GeV which
is now not available anywhere since the closing of the SLAC fixed target program. 

\section{Electron beam of the Electron-Ion Collider}

\begin{figure}[ht!]
	\centering
	\includegraphics[trim = 0 0 0 0, width=0.8\columnwidth, angle = 0.]{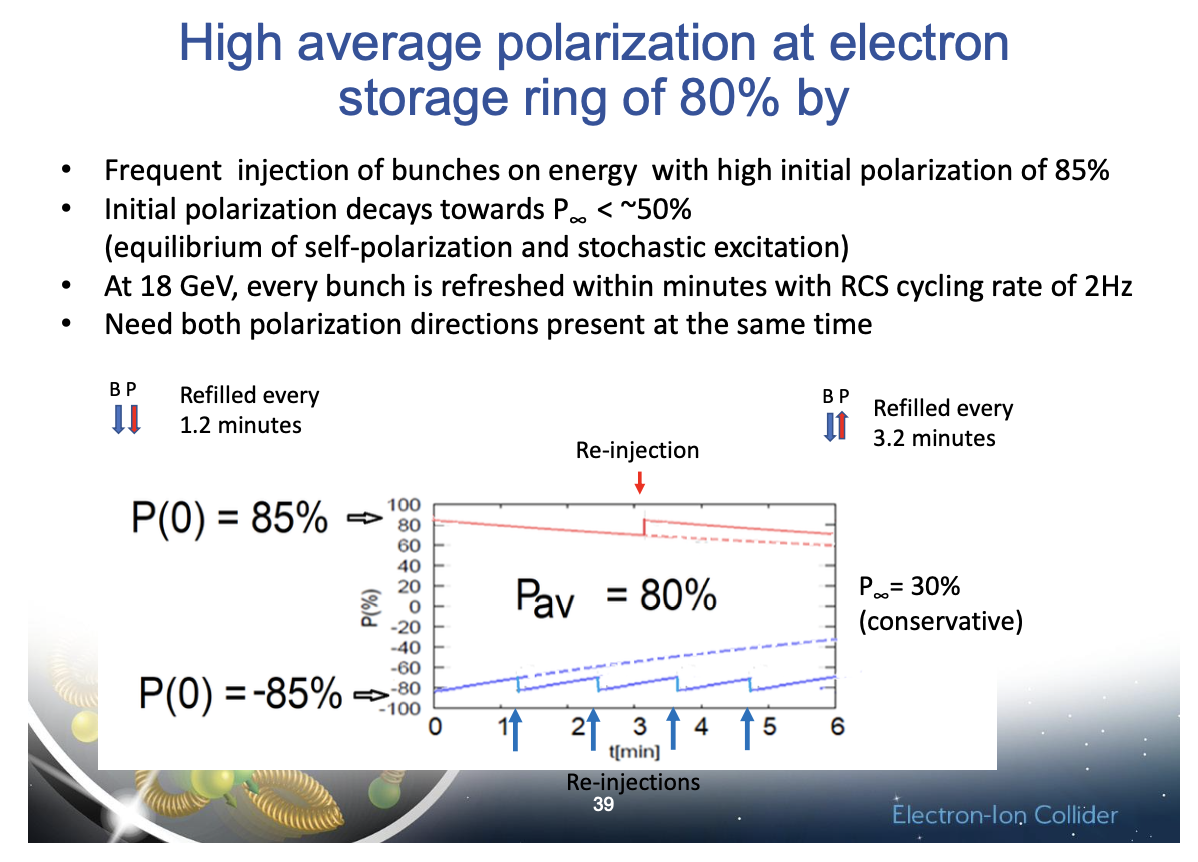}
	\caption{EIC beam polarization and intensity vs. time~\cite{Seryi-21}.}
\label{fig:EIC-beam}
\end{figure}

According to the EIC CDR, the accelerator system includes the Rapid Cycling Synchrotron, which will provide a polarized electron beam into the electron ring where beam polarization as a function of time is shown in Fig.~\ref{fig:EIC-beam} taken from Ref.~\cite{Seryi-21}.
It also allows maintaining high beam current limited primarily by the RF power.

\section{Luminosity and Figure-of-Merit of the internal-target-based experiment at EIC}
For comparison of the Figure-of-Merit in a potential experiment (called for now ``Heracles") at EIC and the one which had been achieved in HERMES  we used information from the EIC accelerator report~\cite{EIC-2021, Seryi-21} and from Ref.~\cite{Spectrometer}.
For HERMES we used average beam current 25~mA, luminosity of $1.2 \times 10^{31}$~electron-nucleon/cm$^2$ (in the case of the deuteron target), 40\% electron beam polarization and 85\% target polarization~\cite{HERMES-cell}.
The target thickness in the case of He-3 was $1\times 10^{15}$~nucleons/cm$^2$ with polarization 54\%, limited by the required lifetime of the beam of 45 hours~\cite{HERMES-He-3}.

At EIC the projected beam current is 227~mA at 18 GeV and 2500~mA at 12 GeV, beam polarization 80\%.
The target thickness usable at EIC can be much higher because of frequent injection of the polarized beam.
It is nine times higher for the triple length of the storage cell (from 40 cm to 120 cm) without significant loss of target polarization.  
For the He-3 target this leads to the EIC FOM with 18 GeV beam of $2 \times10^{33}$
compare to the HERMES FOM $6 \times10^{30}$, both in units electron-nucleons/cm$^2$ with the polarizations in square.
 
One example of a currently needed experiment is connected to the neutron and pion structure function accessible with the soft proton-tagged DIS.
Such an experiment deal with ``free" neutron target.
It was in item III.D of the PEGASYS program, see Fig.~\ref{fig:pegasys}. 
The initial set of data was taken by HERMES during its latest run~\cite{HERMES-recoil} using a soft proton recoil detector 
and by BONUS/CLAS at JLab~\cite{Fenker-08, Christy-19}. 
A specialized setup for such an experiment with luminosity of $10^{36}$~electron-nucleons/cm$^2$ is under development by the TDIS collaboration~\cite{TDIS} for the unpolarized target.
Heracles will allow us to get double polarized data for this process with deuteron target which will allow clean measurement of 
the spin-dependent neutron structure functions.

\section{Layout of the experiment and detector}
\begin{figure}[ht!]
	\centering
	\includegraphics[trim = 0 0 0 0, width=0.7\columnwidth, angle =-0.5]{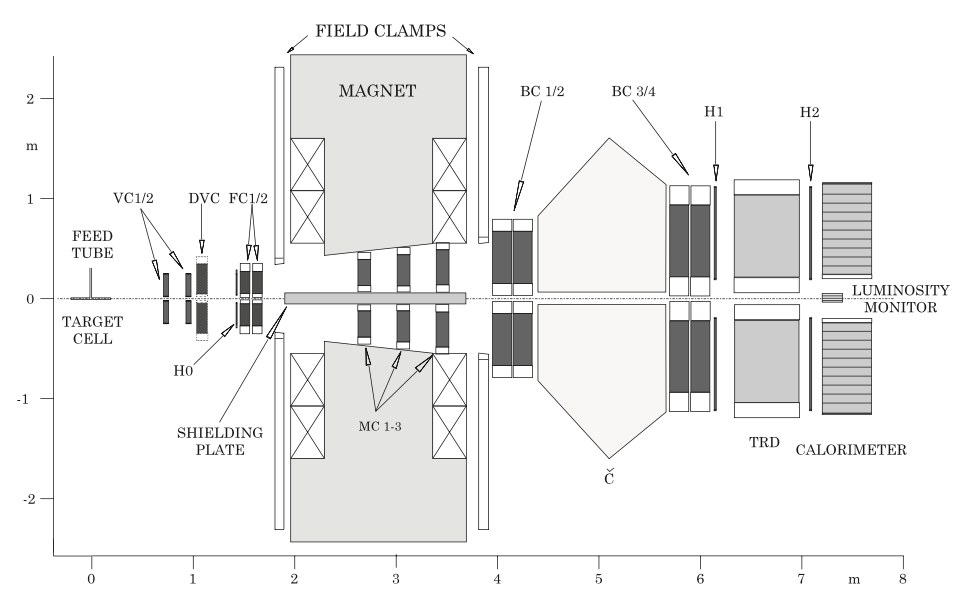}
	\caption{Layout of the HERMES spectrometer~\cite{Spectrometer}.}
\label{fig:spectrometer}
\end{figure}

The layout of the experiment could follow that of HERMES as shown in Fig.~\ref{fig:spectrometer} (or LHCb) with a natural upgrade of the tracking and particle identification detectors, 
as well as the data acquisition electronics, according to the technology developed over the last 30 years.
For example, instead of the wire chambers for tracking Heracles can use modern GEM-based detectors, 
which already operate in the SBS spectrometer at a luminosity of $10^{38}$~electron-nucleons/cm$^2$.

The data will be in the wide range of Q$^2$-x, defined mostly by the beam energy, which is between JLab and HERMES.
However, due to 500+ times higher FOM the data will have 20+ times smaller error bars, which for example
allow dramatic advance in high-x meson production physics.

\section{Conclusion}
The beams of the EIC provide excellent opportunities for double polarized hadron physics experiments, 
especially with a high intensity 18 GeV polarized electron beam.
They could be used for fixed target experiments similar to HERMES with much higher productivity (by a factor of 500+) and
also for the tagged photon beam.\\

\begin{acknowledgments}
I would like to extend thanks to Todd~Averett for fruitful discussions.
This material is based on work supported by the U.S. Department of Energy, Office of Science, 
Office of Nuclear Physics under contract DE-AC05-06OR23177.
\end{acknowledgments}

\bibliography{Heracles_arXiv.bib}

\begin{thebibliography}{26}%
\makeatletter
\providecommand \@ifxundefined [1]{%
 \@ifx{#1\undefined}
}%
\providecommand \@ifnum [1]{%
 \ifnum #1\expandafter \@firstoftwo
 \else \expandafter \@secondoftwo
 \fi
}%
\providecommand \@ifx [1]{%
 \ifx #1\expandafter \@firstoftwo
 \else \expandafter \@secondoftwo
 \fi
}%
\providecommand \natexlab [1]{#1}%
\providecommand \enquote  [1]{``#1''}%
\providecommand \bibnamefont  [1]{#1}%
\providecommand \bibfnamefont [1]{#1}%
\providecommand \citenamefont [1]{#1}%
\providecommand \href@noop [0]{\@secondoftwo}%
\providecommand \href [0]{\begingroup \@sanitize@url \@href}%
\providecommand \@href[1]{\@@startlink{#1}\@@href}%
\providecommand \@@href[1]{\endgroup#1\@@endlink}%
\providecommand \@sanitize@url [0]{\catcode `\\12\catcode `\$12\catcode
  `\&12\catcode `\#12\catcode `\^12\catcode `\_12\catcode `\%12\relax}%
\providecommand \@@startlink[1]{}%
\providecommand \@@endlink[0]{}%
\providecommand \url  [0]{\begingroup\@sanitize@url \@url }%
\providecommand \@url [1]{\endgroup\@href {#1}{\urlprefix }}%
\providecommand \urlprefix  [0]{URL }%
\providecommand \Eprint [0]{\href }%
\providecommand \doibase [0]{https://doi.org/}%
\providecommand \selectlanguage [0]{\@gobble}%
\providecommand \bibinfo  [0]{\@secondoftwo}%
\providecommand \bibfield  [0]{\@secondoftwo}%
\providecommand \translation [1]{[#1]}%
\providecommand \BibitemOpen [0]{}%
\providecommand \bibitemStop [0]{}%
\providecommand \bibitemNoStop [0]{.\EOS\space}%
\providecommand \EOS [0]{\spacefactor3000\relax}%
\providecommand \BibitemShut  [1]{\csname bibitem#1\endcsname}%
\let\auto@bib@innerbib\@empty
\bibitem [{\citenamefont {Tumaikin}(2021)}]{VEP-1}%
  \BibitemOpen
  \bibfield  {author} {\bibinfo {author} {\bibfnamefont {G.}~\bibnamefont
  {Tumaikin}},\ }\bibfield  {title} {\bibinfo {title} {{Development of the
  internal target at VEP-1}},\ }\href
  {https://www.inp.nsk.su/images/Gazeta-2021-N3.pdf} {\bibfield  {journal}
  {\bibinfo  {journal} {{BINP newspaper "Energy-Momentum"}}\ }\textbf {\bibinfo
  {volume} {3}} (\bibinfo {year} {2021})}\BibitemShut {NoStop}%
\bibitem [{\citenamefont {Budker}\ \emph {et~al.}(1967)\citenamefont {Budker},
  \citenamefont {Onuchin}, \citenamefont {Popov},\ and\ \citenamefont
  {Tumaikin}}]{Budker-67}%
  \BibitemOpen
  \bibfield  {author} {\bibinfo {author} {\bibfnamefont {G.}~\bibnamefont
  {Budker}}, \bibinfo {author} {\bibfnamefont {A.}~\bibnamefont {Onuchin}},
  \bibinfo {author} {\bibfnamefont {S.}~\bibnamefont {Popov}},\ and\ \bibinfo
  {author} {\bibfnamefont {G.}~\bibnamefont {Tumaikin}},\ }\bibfield  {title}
  {\bibinfo {title} {Experiments with a target in the electron storage ring},\
  }\href {https://link.springer.com/journal/11450} {\bibfield  {journal}
  {\bibinfo  {journal} {Soviet journal of nuclear physics}\ }\textbf {\bibinfo
  {volume} {6}},\ \bibinfo {pages} {775} (\bibinfo {year} {1967})}\BibitemShut
  {NoStop}%
\bibitem [{\citenamefont {Zelevinskii}\ \emph {et~al.}(1969)\citenamefont
  {Zelevinskii}, \citenamefont {Nikolenko}, \citenamefont {Popov},\ and\
  \citenamefont {Tumaikin}}]{Zelevinski-69}%
  \BibitemOpen
  \bibfield  {author} {\bibinfo {author} {\bibfnamefont {V.}~\bibnamefont
  {Zelevinskii}}, \bibinfo {author} {\bibfnamefont {D.}~\bibnamefont
  {Nikolenko}}, \bibinfo {author} {\bibfnamefont {S.}~\bibnamefont {Popov}},\
  and\ \bibinfo {author} {\bibfnamefont {G.}~\bibnamefont {Tumaikin}},\
  }\bibfield  {title} {\bibinfo {title} {Use of the electron storage ring for
  experiments on the electron excitation of nuclei.},\ }\href@noop {}
  {\bibfield  {journal} {\bibinfo  {journal} {Izvestiya Akademii Nauk USSR,
  Seriya Fizicheskaya}\ }\textbf {\bibinfo {volume} {33}},\ \bibinfo {pages}
  {686} (\bibinfo {year} {1969})}\BibitemShut {NoStop}%
\bibitem [{\citenamefont {Belyaev}\ \emph {et~al.}(1970)\citenamefont
  {Belyaev}, \citenamefont {Budker},\ and\ \citenamefont {Popov}}]{Belyaev-70}%
  \BibitemOpen
  \bibfield  {author} {\bibinfo {author} {\bibfnamefont {S.}~\bibnamefont
  {Belyaev}}, \bibinfo {author} {\bibfnamefont {G.}~\bibnamefont {Budker}},\
  and\ \bibinfo {author} {\bibfnamefont {S.}~\bibnamefont {Popov}},\ }\bibfield
   {title} {\bibinfo {title} {The possibility of using storage rings with
  internal thin targets},\ }in\ \href@noop {} {\emph {\bibinfo {booktitle}
  {High-Energy Physics and Nuclear Structure: Proceedings of the Third
  International Conference on High Energy Physics and Nuclear Structure
  sponsored by the International Union of Pure and Applied Physics, held at
  Columbia University, New York City, September 8--12, 1969}}}\ (\bibinfo
  {organization} {Springer},\ \bibinfo {year} {1970})\ pp.\ \bibinfo {pages}
  {606--609}\BibitemShut {NoStop}%
\bibitem [{\citenamefont {Popov}(1983)}]{Popov-82}%
  \BibitemOpen
  \bibfield  {author} {\bibinfo {author} {\bibfnamefont {S.}~\bibnamefont
  {Popov}},\ }\href {https://books.google.com/books?id=Dc7vAAAAMAAJ} {\bibinfo
  {title} {Experiments with a target in an electron storage ring}} (\bibinfo
  {year} {1983})\BibitemShut {NoStop}%
\bibitem [{\citenamefont {Dmitriev}\ \emph {et~al.}(1985)\citenamefont
  {Dmitriev} \emph {et~al.}}]{Dmitriev-85}%
  \BibitemOpen
  \bibfield  {author} {\bibinfo {author} {\bibfnamefont {V.}~\bibnamefont
  {Dmitriev}} \emph {et~al.},\ }\bibfield  {title} {\bibinfo {title} {First
  measurement of the asymmetry in electron scattering by a jet target of
  polarized deuterium atoms},\ }\href
  {https://doi.org/https://doi.org/10.1016/0370-2693(85)91534-5} {\bibfield
  {journal} {\bibinfo  {journal} {Physics Letters B}\ }\textbf {\bibinfo
  {volume} {157}},\ \bibinfo {pages} {143} (\bibinfo {year}
  {1985})}\BibitemShut {NoStop}%
\bibitem [{\citenamefont {Popov}(1987)}]{Popov-87}%
  \BibitemOpen
  \bibfield  {author} {\bibinfo {author} {\bibfnamefont {S.}~\bibnamefont
  {Popov}},\ }\bibfield  {title} {\bibinfo {title} {{Experiments with an
  internal target in the electron storage ring}},\ }in\ \href@noop {} {\emph
  {\bibinfo {booktitle} {{International Symposium on Modern Developments in
  Nuclear Physics 27 June-1 July 1987. Novosibirsk, USSR}}}}\ (\bibinfo {year}
  {1987})\BibitemShut {NoStop}%
\bibitem [{\citenamefont {Baturin}\ \emph {et~al.}(1989)\citenamefont {Baturin}
  \emph {et~al.}}]{NEP}%
  \BibitemOpen
  \bibfield  {author} {\bibinfo {author} {\bibfnamefont {P.}~\bibnamefont
  {Baturin}} \emph {et~al.},\ }\bibfield  {title} {\bibinfo {title}
  {{Specialized storage ring with longitudinally polarized electrons for
  internal target experiments}},\ }\href {https://doi.org/10.1063/1.38324}
  {\bibfield  {journal} {\bibinfo  {journal} {AIP Conference Proceedings}\
  }\textbf {\bibinfo {volume} {187}},\ \bibinfo {pages} {1028} (\bibinfo {year}
  {1989})}\BibitemShut {NoStop}%
\bibitem [{\citenamefont {Gilman}\ \emph {et~al.}(1990)\citenamefont {Gilman}
  \emph {et~al.}}]{Gilman-90}%
  \BibitemOpen
  \bibfield  {author} {\bibinfo {author} {\bibfnamefont {R.}~\bibnamefont
  {Gilman}} \emph {et~al.},\ }\bibfield  {title} {\bibinfo {title} {Measurement
  of tensor analyzing power in electron-deuteron elastic scattering},\ }\href
  {https://doi.org/10.1103/PhysRevLett.65.1733} {\bibfield  {journal} {\bibinfo
   {journal} {Phys. Rev. Lett.}\ }\textbf {\bibinfo {volume} {65}},\ \bibinfo
  {pages} {1733} (\bibinfo {year} {1990})}\BibitemShut {NoStop}%
\bibitem [{\citenamefont {Milner}(1985)}]{Milner-84}%
  \BibitemOpen
  \bibfield  {author} {\bibinfo {author} {\bibfnamefont {R.}~\bibnamefont
  {Milner}},\ }\href {https://doi.org/10.1063/1.35342} {\bibinfo {title}
  {Investigation of the internal spin structure of the neutron by deep
  inelastic scattering of longitudinally polarized electrons from a polarized
  {H}e-3 target}} (\bibinfo {year} {1985})\BibitemShut {NoStop}%
\bibitem [{\citenamefont {{Van Bibber}}(1989)}]{VANBIBBER}%
  \BibitemOpen
  \bibfield  {author} {\bibinfo {author} {\bibfnamefont {K.}~\bibnamefont {{Van
  Bibber}}},\ }\bibfield  {title} {\bibinfo {title} {{PEGASYS — a proposed
  internal target spectrometer facility for the PEP storage ring}},\ }\href
  {https://doi.org/https://doi.org/10.1016/0168-583X(89)91015-X} {\bibfield
  {journal} {\bibinfo  {journal} {Nuclear Instruments and Methods in Physics
  Research Section B: Beam Interactions with Materials and Atoms}\ }\textbf
  {\bibinfo {volume} {40-41}},\ \bibinfo {pages} {436} (\bibinfo {year}
  {1989})}\BibitemShut {NoStop}%
\bibitem [{\citenamefont {McKeown}(1987)}]{McKeown-87}%
  \BibitemOpen
  \bibfield  {author} {\bibinfo {author} {\bibfnamefont {R.}~\bibnamefont
  {McKeown}},\ }\bibfield  {title} {\bibinfo {title} {Possibilities for
  polarized internal targets},\ }in\ \href
  {https://www.osti.gov/biblio/6993778} {\emph {\bibinfo {booktitle} {{NPAS
  Workshop on Electronuclear Physics with Internal Targets}}}}\ (\bibinfo
  {year} {1987})\ pp.\ \bibinfo {pages} {99--102}\BibitemShut {NoStop}%
\bibitem [{\citenamefont {Milner}(1987)}]{Milner-87}%
  \BibitemOpen
  \bibfield  {author} {\bibinfo {author} {\bibfnamefont {R.}~\bibnamefont
  {Milner}},\ }\bibfield  {title} {\bibinfo {title} {Electromagnetic physics
  with a polarized he-3 internal target},\ }in\ \href
  {https://www.osti.gov/biblio/6993778} {\emph {\bibinfo {booktitle} {{NPAS
  Workshop on Electronuclear Physics with Internal Targets}}}}\ (\bibinfo
  {year} {1987})\ pp.\ \bibinfo {pages} {195--199}\BibitemShut {NoStop}%
\bibitem [{\citenamefont {Milner}\ and\ \citenamefont
  {Steffens}(2021)}]{Milner-2021}%
  \BibitemOpen
  \bibfield  {author} {\bibinfo {author} {\bibfnamefont {R.}~\bibnamefont
  {Milner}}\ and\ \bibinfo {author} {\bibfnamefont {E.}~\bibnamefont
  {Steffens}},\ }\href {https://doi.org/10.1142/11692} {\emph {\bibinfo {title}
  {{The HERMES Experiment}: {A Personal Story}}}}\ (\bibinfo  {publisher}
  {World Scientific},\ \bibinfo {year} {2021})\BibitemShut {NoStop}%
\bibitem [{\citenamefont {{PEGASYS-MarkII}}(1990)}]{PEGASYS}%
  \BibitemOpen
  \bibfield  {author} {\bibinfo {author} {\bibnamefont {{PEGASYS-MarkII}}},\
  }\href
  {https://www.slac.stanford.edu/pubs/slacreports/reports15/slac-r-377.pdf}
  {\bibinfo {title} {{A Program of Internal Target Physics Using the Mark-II
  Detector at the PEP Storage Ring}}} (\bibinfo {year} {1990})\BibitemShut
  {NoStop}%
\bibitem [{\citenamefont {{Abdul Khalek}}\ \emph {et~al.}(2022)\citenamefont
  {{Abdul Khalek}} \emph {et~al.}}]{EIC-Yellow}%
  \BibitemOpen
  \bibfield  {author} {\bibinfo {author} {\bibfnamefont {R.}~\bibnamefont
  {{Abdul Khalek}}} \emph {et~al.},\ }\bibfield  {title} {\bibinfo {title}
  {{Science Requirements and Detector Concepts for the Electron-Ion Collider:
  EIC Yellow Report}},\ }\href
  {https://doi.org/https://doi.org/10.1016/j.nuclphysa.2022.122447} {\bibfield
  {journal} {\bibinfo  {journal} {Nuclear Physics A}\ }\textbf {\bibinfo
  {volume} {1026}},\ \bibinfo {pages} {122447} (\bibinfo {year} {2022})},\
  \bibinfo {note} {page 650, section 12.5}\BibitemShut {NoStop}%
\bibitem [{\citenamefont {Wojtsekhowski}(2023)}]{BW-23}%
  \BibitemOpen
  \bibfield  {author} {\bibinfo {author} {\bibfnamefont {B.}~\bibnamefont
  {Wojtsekhowski}},\ }\href
  {https://indico.cern.ch/event/1199314/contributions/5193183/attachments/2619466/4528559/Wojtsekhowski-DIS2023.pdf}
  {\bibinfo {title} {{Polarized internal target experiments with the EIC
  beams}}} (\bibinfo {year} {2023})\BibitemShut {NoStop}%
\bibitem [{\citenamefont {Seryi}(2021)}]{Seryi-21}%
  \BibitemOpen
  \bibfield  {author} {\bibinfo {author} {\bibfnamefont {A.}~\bibnamefont
  {Seryi}},\ }\href
  {https://indico.jlab.org/event/438/contributions/8833/attachments/7263/10020/EICUG-July-29-2021-Seryi-r2.pdf}
  {\bibinfo {title} {The electron-ion collider – accelerator design
  overview}} (\bibinfo {year} {2021})\BibitemShut {NoStop}%
\bibitem [{\citenamefont {Willeke}\ and\ \citenamefont
  {Beebe-Wang}(2021)}]{EIC-2021}%
  \BibitemOpen
  \bibfield  {author} {\bibinfo {author} {\bibfnamefont {F.}~\bibnamefont
  {Willeke}}\ and\ \bibinfo {author} {\bibfnamefont {J.}~\bibnamefont
  {Beebe-Wang}},\ }\bibfield  {title} {\bibinfo {title} {Electron ion collider
  conceptual design report 2021},\ }\bibfield  {journal} {\bibinfo  {journal}
  {{BNL/EIC web page}}\ }\href {https://doi.org/10.2172/1765663}
  {10.2172/1765663} (\bibinfo {year} {2021})\BibitemShut {NoStop}%
\bibitem [{\citenamefont {Ackerstaff}\ \emph {et~al.}(1998)\citenamefont
  {Ackerstaff} \emph {et~al.}}]{Spectrometer}%
  \BibitemOpen
  \bibfield  {author} {\bibinfo {author} {\bibfnamefont {K.}~\bibnamefont
  {Ackerstaff}} \emph {et~al.},\ }\bibfield  {title} {\bibinfo {title} {The
  hermes spectrometer},\ }\href
  {https://doi.org/https://doi.org/10.1016/S0168-9002(98)00769-4} {\bibfield
  {journal} {\bibinfo  {journal} {Nuclear Instruments and Methods in Physics
  Research Section A: Accelerators, Spectrometers, Detectors and Associated
  Equipment}\ }\textbf {\bibinfo {volume} {417}},\ \bibinfo {pages} {230}
  (\bibinfo {year} {1998})}\BibitemShut {NoStop}%
\bibitem [{\citenamefont {Baumgarten}\ \emph {et~al.}(2003)\citenamefont
  {Baumgarten} \emph {et~al.}}]{HERMES-cell}%
  \BibitemOpen
  \bibfield  {author} {\bibinfo {author} {\bibfnamefont {C.}~\bibnamefont
  {Baumgarten}} \emph {et~al.},\ }\bibfield  {title} {\bibinfo {title} {{The
  storage cell of the polarized H/D internal gas target of the HERMES
  experiment at HERA}},\ }\href {https://doi.org/10.1016/S0168-9002(02)01752-7}
  {\bibfield  {journal} {\bibinfo  {journal} {Nucl. Instrum. Meth. A}\ }\textbf
  {\bibinfo {volume} {496}},\ \bibinfo {pages} {277} (\bibinfo {year}
  {2003})}\BibitemShut {NoStop}%
\bibitem [{\citenamefont {DeSchepper}\ \emph {et~al.}(1998)\citenamefont
  {DeSchepper} \emph {et~al.}}]{HERMES-He-3}%
  \BibitemOpen
  \bibfield  {author} {\bibinfo {author} {\bibfnamefont {D.}~\bibnamefont
  {DeSchepper}} \emph {et~al.},\ }\bibfield  {title} {\bibinfo {title} {{The
  HERMES polarized He-3 internal gas target}},\ }\href
  {https://doi.org/10.1016/S0168-9002(98)00901-2} {\bibfield  {journal}
  {\bibinfo  {journal} {Nucl. Instrum. Meth. A}\ }\textbf {\bibinfo {volume}
  {419}},\ \bibinfo {pages} {16} (\bibinfo {year} {1998})}\BibitemShut
  {NoStop}%
\bibitem [{\citenamefont {Airapetian}\ \emph {et~al.}(2013)\citenamefont
  {Airapetian} \emph {et~al.}}]{HERMES-recoil}%
  \BibitemOpen
  \bibfield  {author} {\bibinfo {author} {\bibfnamefont {A.}~\bibnamefont
  {Airapetian}} \emph {et~al.},\ }\bibfield  {title} {\bibinfo {title} {The
  hermes recoil detector},\ }\href
  {https://doi.org/10.1088/1748-0221/8/05/P05012} {\bibfield  {journal}
  {\bibinfo  {journal} {Journal of Instrumentation}\ }\textbf {\bibinfo
  {volume} {8}}\bibinfo  {number} { (05)},\ \bibinfo {pages}
  {P05012}}\BibitemShut {NoStop}%
\bibitem [{\citenamefont {Fenker}\ \emph {et~al.}(2008)\citenamefont {Fenker}
  \emph {et~al.}}]{Fenker-08}%
  \BibitemOpen
\bibfield  {number} {  }\bibfield  {author} {\bibinfo {author} {\bibfnamefont
  {H.}~\bibnamefont {Fenker}} \emph {et~al.},\ }\bibfield  {title} {\bibinfo
  {title} {Bonus: Development and use of a radial tpc using cylindrical gems},\
  }\href {https://doi.org/https://doi.org/10.1016/j.nima.2008.04.047}
  {\bibfield  {journal} {\bibinfo  {journal} {Nuclear Instruments and Methods
  in Physics Research Section A: Accelerators, Spectrometers, Detectors and
  Associated Equipment}\ }\textbf {\bibinfo {volume} {592}},\ \bibinfo {pages}
  {273} (\bibinfo {year} {2008})}\BibitemShut {NoStop}%
\bibitem [{\citenamefont {Christy}(2019)}]{Christy-19}%
  \BibitemOpen
  \bibfield  {author} {\bibinfo {author} {\bibfnamefont {E.}~\bibnamefont
  {Christy}},\ }\href
  {https://indico.cern.ch/event/799284/contributions/3478975/attachments/1894493/3125191/B01-christy_hix2019.pdf}
  {\bibinfo {title} {{JLab Hall B BONUS Experiment neutron structure}}}
  (\bibinfo {year} {2019})\BibitemShut {NoStop}%
\bibitem [{\citenamefont {Keppel}\ \emph {et~al.}(2014)\citenamefont {Keppel}
  \emph {et~al.}}]{TDIS}%
  \BibitemOpen
  \bibfield  {author} {\bibinfo {author} {\bibfnamefont {C.}~\bibnamefont
  {Keppel}} \emph {et~al.},\ }\href
  {https://www.jlab.org/exp_prog/proposals/14/PR12-14-010.pdf} {\bibinfo
  {title} {{Measurement of Tagged Deep Inelastic Scattering (TDIS)}}} (\bibinfo
  {year} {2014})\BibitemShut {NoStop}%
\end{thebibliography}%

\end{document}